\documentclass[preprint,12pt]{elsarticle}

\usepackage[T1]{fontenc}
\usepackage[utf8]{inputenc}
\usepackage{textcomp}
\usepackage{graphicx}
\usepackage{amsmath,amssymb}
\usepackage{hyperref}
\usepackage{url}
\journal{Nuclear Materials and Energy}
\biboptions{sort&compress}

\begin{document}
\begin{frontmatter}

\title{Analysis of Solid Boron Injection for ITER}

\author[]{F. Effenberg\textsuperscript{a*\&}}
\author[]{J. A. Snipes\textsuperscript{a,\&}}
\author[]{F. Nespoli\textsuperscript{a}}
\author[]{R. Lunsford\textsuperscript{a}}
\author[]{J. van Blarcum\textsuperscript{b}}
\author[]{M. Jia\textsuperscript{c}}
\author[]{H. Frerichs\textsuperscript{d}}
\author[]{A. Bortolon\textsuperscript{a}}
\author[]{K. Schmid\textsuperscript{e}}
\author[]{T. Wauters\textsuperscript{b}}

\nonumnote{\textsuperscript{*} Corresponding author \href{mailto:feffenbe@pppl.gov}{feffenbe@pppl.gov} (F. Effenberg)}
\nonumnote{\textsuperscript{\&} Equal first authors.}

\address[a]{Princeton Plasma Physics Laboratory, 100 Stellarator Road, Princeton, NJ 08540 USA}
\address[b]{ITER Organization, Route de Vinon-sur-Verdon, CS 90 046, 13067 St Paul Lez Durance Cedex, France}
\address[c]{Institute of Plasma Physics, Chinese Academy of Sciences, P.O. Box 1126, Hefei 230031, China}
\address[d]{Department of Nuclear Engineering and Engineering Physics, University of Wisconsin, Madison, WI 53706, USA}
\address[e]{Max-Planck-Institut f\"ur Plasmaphysik, Boltzmannstra{\ss}e 2, D-85748 Garching b. M\"unchen, Germany}

\begin{abstract}
Since ITER will now operate with a full tungsten divertor and first wall \cite{ref1}, the ITER Organization is considering boron wall conditioning schemes to getter oxygen and reduce tungsten sputtering to avoid excessive radiation to ensure that ITER can reach and sustain high- performance plasma operation. ITER has adopted Glow Discharge Boronization (GDB) \cite{ref2} in deuterated diborane (B\textsubscript{2}D\textsubscript{6}) as its main technique, which can only be performed in the absence of magnetic fields, but has also requested a study to determine how effective solid boron injection (SBI), which can be performed during plasma operation, using either an Impurity Powder Dropper (IPD) \cite{ref3} or a boron pellet injector \cite{ref4}, might be as a risk mitigation to achieving Q=10. Solid impurity injection has been used for decades as a wall conditioning technique on numerous tokamaks and stellarators around the world \cite{ref5,ref6,ref7,ref8,ref9,ref10,ref11,ref12,ref13}. The study presented here begins to assess SBI on ITER by first quantifying the boron source term across a particle size scan from 1\textmu m to 5 mm diameter and an injection velocity scan from 5 -- 200 m/s for several ITER plasma conditions: 1) Start of Research Operations (SRO) diverted L-mode, 2) Q=10 H-mode, and 3) low plasma current and low power limiter operation. Modeling of the ablation, transport, erosion and redeposition of solid boron is being carried out with the Dust Injection Simulator (DIS) \cite{ref14}, the Neutral Gas Shielding (NGS) \cite{ref15,ref16,ref17}, EMC3-EIRENE \cite{ref18,ref19} and WallDYN3D \cite{ref20} models. In the first steps presented here, the DIS and NGS models are compared for calculating the boron source term that is then used as input to EMC3-EIRENE and will later be used with WallDYN3D.
\end{abstract}


\end{frontmatter}

\section{Introduction}
\label{sec:intro}

The decision on ITER to avoid beryllium as a first wall material and instead operate with a full tungsten first wall and divertor \cite{ref1} has led the ITER team to adopt Glow Discharge Boronization (GDB) \cite{ref2} in deuterated diborane (B\textsubscript{2}D\textsubscript{6}) to getter oxygen to reduce the sputtering of tungsten into the plasma. In addition, as risk mitigation to achieving and sustaining high performance plasma operation to reach their Q=10 goal, ITER is also considering adding an actuator for real-time solid boron injection (SBI) during plasma operation with either an Impurity Powder Dropper (IPD) \cite{ref3} or an impurity pellet injector \cite{ref4} and has asked for an assessment of the effectiveness of such a system, particularly for oxygen gettering to reduce tungsten sputtering, under ITER plasma conditions. This is particularly important in a machine with superconducting coils such as ITER since GDB can only be performed with the magnets off during maintenance periods, whereas SBI is performed during plasma operation. Solid impurity injection using pellets or powders of various materials such as lithium, boron, and boron nitride has been used for decades as a wall conditioning technique on numerous tokamaks and stellarators around the world \cite{ref5,ref6,ref7,ref8,ref9,ref10,ref11,ref12,ref13}, which even led to the highest DT fusion performance in TFTR \cite{ref21} and to the longest high performance plasmas in EAST exceeding 1000 s duration \cite{ref22,ref23}.

This assessment of SBI on ITER has begun by modeling the boron source for a range of injected particle sizes from 1 \textmu m -- 5 mm diameter, with DIS applied up to 1.5 mm and NGS up to 5 mm, and particle velocities from 5 -- 200 m/s to understand how far the particles will penetrate into the plasma under ITER conditions. These particle size and injection velocity ranges are chosen for practical considerations to vary the expected penetration depth across the LCFS and subsequent plasma transport to plasma- facing surfaces. Three ITER scenarios have been modeled initially in a 20 degree toroidal sector for 1) a diverted L-mode scenario with 40 MW of input power and edge electron density of 1.2 x 10\textsuperscript{19} m\textsuperscript{-3}, 2) a Q=10 H-mode scenario with 100 MW of input power and edge electron density of 4 x 10\textsuperscript{19} m\textsuperscript{-3}, and 3) a limiter start-up scenario with 5 MW of input power and edge electron density of 1 x 10\textsuperscript{19} m\textsuperscript{-3}. The Dust Injection Simulator (DIS) code \cite{ref14} was used to model powder injection with particle sizes from 1 \textmu m -- 1.5 mm diameter and injection velocities from 5 -- 200 m/s. The Neutral Gas Shielding (NGS) model \cite{ref15,ref16,ref17} was used to model particles from 1 \textmu m -- 5 mm in diameter with pellet injection velocities from 5 -- 200 m/s.

\begin{figure}[htbp]
\centering
\includegraphics[width=0.40\linewidth]{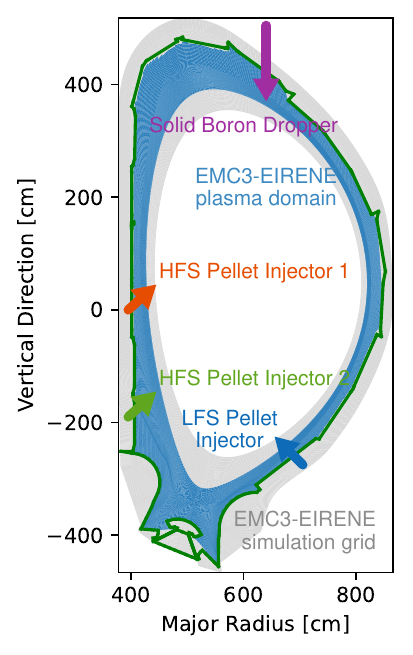}
\caption{ITER poloidal cross-section showing the modeled solid-boron injection locations and EMC3-EIRENE computational domains. The magenta arrow indicates the upper-port solid boron dropper, the orange and green arrows indicate the two high-field-side (HFS) pellet injector locations, and the blue arrow indicates the low-field-side (LFS) pellet injector location; arrows denote the corresponding injection directions. The light-blue region marks the EMC3-EIRENE plasma domain, while the gray region indicates the extended simulation grid used to resolve plasma transport near the first wall.}
\label{fig:1}
\end{figure}

Next, the calculated boron source was input into EMC3-EIRENE \cite{ref18} to model the boron charge state distribution and transport in the plasma for later use in the WallDYN3D \cite{ref20,ref24} code to model the transport, erosion, and redeposition of boron around the machine to understand how much the initial penetration and ablation of the particles determines their effectiveness in coating the tungsten surfaces in ITER.

\section{Quantification of the boron source term under different ITER background plasma conditions}
\label{sec:source}

The boron source term that is required input for subsequent plasma transport modeling is the quantity of boron injected in atoms per cubic meter together with its injection locations in the plasma and injection velocities. Three ITER plasma scenarios have been modeled with SBI to understand how the depth of penetration and material ablation depend on plasma parameters, particle size, and injection velocity. The plasma profiles of density, temperature, current, and edge power flux are input to determine the ablation rate of the injected particles, and they are calculated for the ITER scenarios produced by the JINTRAC integrated modeling suite, which couples core transport, scrape-off layer (SOL) and divertor physics, magnetic equilibrium, neutral dynamics, and auxiliary heating into a single framework \cite{ref25}. See reference \cite{ref25} for details of the transport models used in the JINTRAC suite. The ITER scenarios modeled include 1) a moderate auxiliary heating power scenario with 40 MW of input power in L-mode for the Start of Research Operations (SRO) phase; 2) a Q=10 15 MA H-mode scenario with 100 MW of input power and a neon concentration of 1.2\% using the same magnetic equilibrium as in the L-mode case; and 3) a low current and low input power (5 MW) limiter start-up scenario.

\begin{figure}[htbp]
\centering
\includegraphics[width=\linewidth]{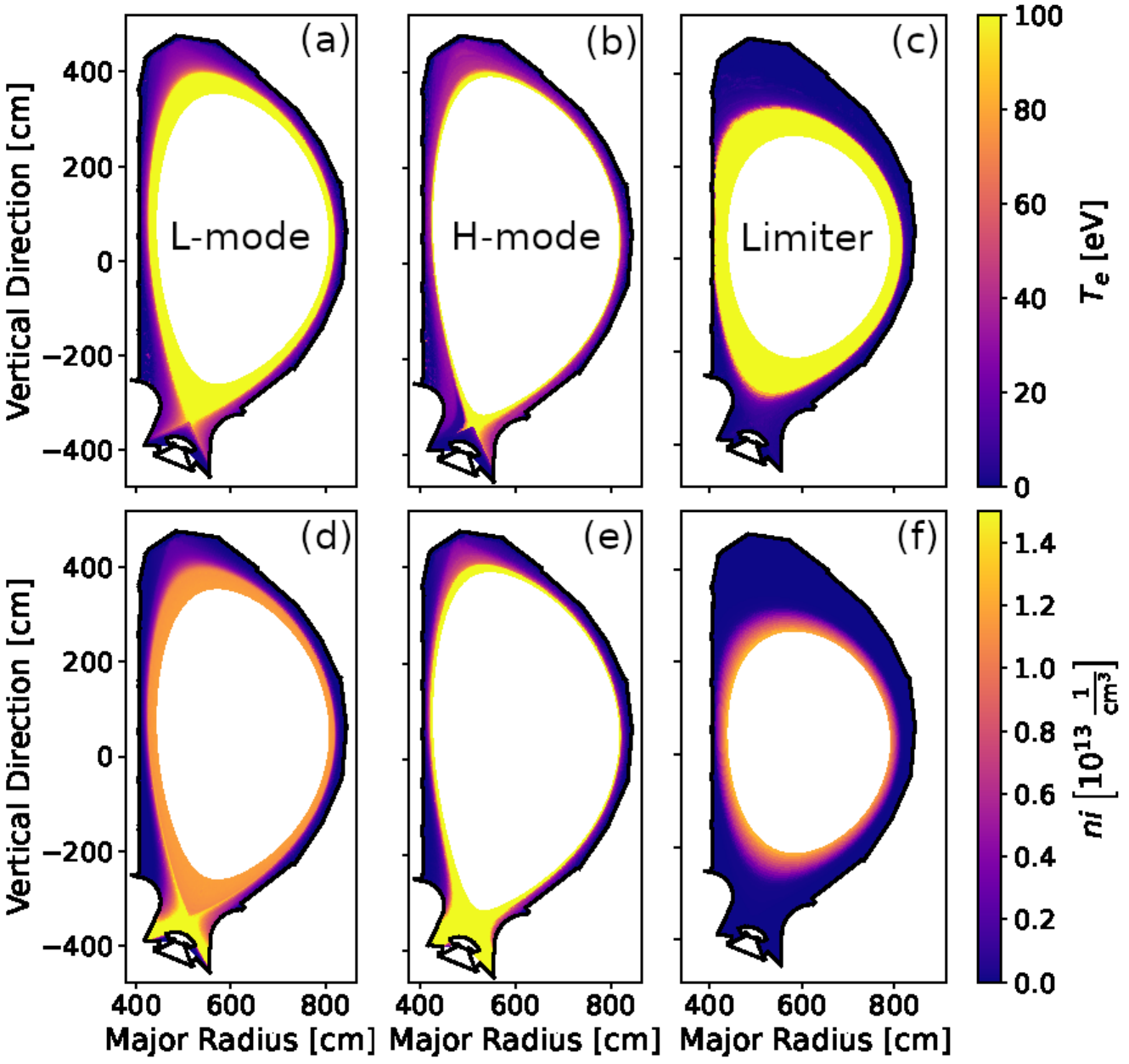}
\caption{Background plasma conditions used for the solid-boron ablation and source-term calculations in the three ITER scenarios. Electron temperature T\textsubscript{e} is shown for (a) SRO diverted L-mode, (b) Q = 10 H-mode, and (c) low-current limiter operation; the corresponding plasma ion density n\textsubscript{i} is shown in (d)--(f), respectively.}
\label{fig:2}
\end{figure}

The boron is modeled as injected either by pellets through one of the existing ITER inboard or outboard pellet guide tubes or by an IPD through an upper port with an assumed entry point of R=6.4 m, Z=4.25 m (Figure \ref{fig:1}). The DIS model is particularly applicable to small boron granules to model transport, deposition, and spatial dispersion in the plasma edge. The NGS model is more applicable to large boron pellets, taking into account the shielding of a neutral gas cloud that surrounds larger pellets during the ablation process. The two models are compared over their overlapping particle-size and velocity ranges to assess how their different treatments of particle heating and neutral-gas shielding affect the predicted penetration and boron source location. Neither model is assumed to be generally more accurate; the comparison tests the sensitivity of the predicted source term to their different physical assumptions. A nominal drop velocity of 5 m/s is considered as the baseline for gravitational injection from an upper ITER port, while higher velocities up to 200 m/s are modeled for larger pellet injection to assess the penetration dependence and operational flexibility. The three ITER scenarios were implemented into EMC3-EIRENE coupled to DIS to calculate the boron source in a 20$^\circ$ toroidal plasma domain that will later be expanded to a full 360$^\circ$ domain for the subsequent WallDyn3D calculations of erosion and redeposition. The particulate source is located toroidally at $\phi$=9.99$^\circ$, toroidally centered within the grid domain, which extends from 0-20$^\circ$. Figure \ref{fig:2} shows the plasma background electron temperature and ion density calculated with EMC3-EIRENE for the three ITER scenarios.

The DIS model is well suited for smaller particles injected gravitationally at low injection velocities. It includes plasma drag, gravity, thermal conduction, surface energy, and current balance (charge). Heating and ablation of particles are mainly determined by the local plasma temperature and density, while also influenced by the plasma flow. For the comparison between DIS and NGS, the plasma flow was turned off in DIS. The NGS model, on the other hand, is well suited for injecting larger particles at higher velocities. The heat flux on the injected particle is shielded by an empirically determined factor $\eta$ = 0.7 \cite{ref17}. Once a finite incident heat flux is encountered along the particle trajectory, the boron particle surface temperature is assumed to remain constant at the boron evaporation temperature. As a result, if shielding is turned off in the NGS model by setting $\eta$ = 1, penetration is greatly reduced to the very edge of the plasma. Reference \cite{ref26} describes a scan of $\eta$ from 0.1 to 1 under ITER conditions. In contrast, the DIS model calculates the time-dependent thermal evolution of the particle, including the effect of finite thermal inertia.

\section{Boron particle size scan at a fixed gravitational injection velocity into the ITER plasma scenarios}
\label{sec:size}

\begin{figure}[htbp]
\centering
\includegraphics[width=\linewidth]{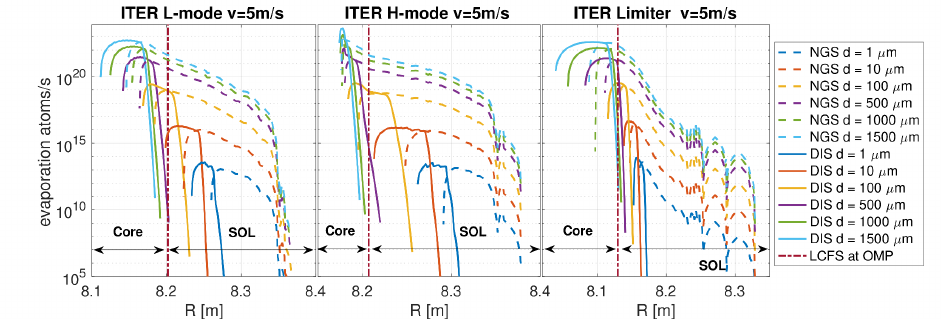}
\caption{Radial boron evaporation-rate profiles predicted by the DIS (solid lines) and NGS (dashed lines) models for particle sizes of 1--1500 \textmu m injected at 5 m/s in the three ITER scenarios: SRO L-mode, Q=10 H-mode, and low-power limiter plasma. Vertical dashed lines indicate the LCFS position for each scenario.}
\label{fig:3}
\end{figure}

To assess the dependence of penetration depth on particle size, a comparison was made between the DIS and NGS models for a fixed velocity of 5 m/s assumed injected gravitationally from an IPD in an upper port. The boron particle size was scanned from 1 \textmu m -- 1500 \textmu m diameter. Figure \ref{fig:3} shows the ablated material in atoms per second versus the outboard major radius at the vertical position of the plasma magnetic axis for the three ITER plasma scenarios.

The vertical lines in Fig. \ref{fig:3} show the locations of the last closed flux surface (LCFS) for each scenario. The DIS simulations (solid lines) begin to ablate much deeper into the plasma and finish ablating somewhat deeper into the plasma than the NGS simulations (dashed lines), especially for the L-mode and limiter cases. This result, at first, seems counterintuitive since the neutral gas shielding effects in the NGS model are not explicitly included in the DIS model. However, the assumption in the NGS model that the particle surface temperature is always at the boron evaporation temperature as soon as it enters the SOL plasma explains why the NGS simulations begin to ablate far out in the SOL, whereas the DIS model includes thermal conduction of the particle in the background plasma temperature and density and the resulting surface energy balance that takes time for the particle to reach evaporation temperature. As a result, the particles in DIS do not begin or finish ablating until deeper into the plasma, though in the Q=10 H-mode case, the very steep edge pedestal ablates even the largest particles rather quickly once they are within the LCFS.

These results show that the low current and input power limiter case provides the weakest plasma screening of the three scenarios considered. At an injection velocity of 5 m/s, the smallest particles remain predominantly in the SOL, while particles of approximately 100 \textmu m and larger increasingly penetrate across the LCFS, with larger particles ablating progressively deeper in the confined plasma region.

In the moderate power L-mode SRO scenario, particles with diameters larger than about 100 \textmu m diameter are expected to penetrate beyond the LCFS at this low injection velocity. Given the density and stored energy of ITER plasmas, 100 \textmu m diameter particles should not substantially perturb the plasma, and so their use would be warranted. Indeed, very fine powders such as the 1 \textmu m diameter particles simulated might not be suitable for ITER since they would be subject to drag and electrostatic forces that may complicate injection. Such fine particles are only modeled here for completeness.

In the high-power, Q=10, 15 MA scenario simulated, only larger particles with diameters greater than 100 \textmu m would be expected to deposit most of their mass inside the LCFS at this low injection velocity. This result suggests that injection into such high-performance plasmas would require larger pellets to ensure penetration into the confined plasma region. An injection velocity scan is described in section 4.

\begin{figure}[htbp]
\centering
\includegraphics[width=0.65\linewidth]{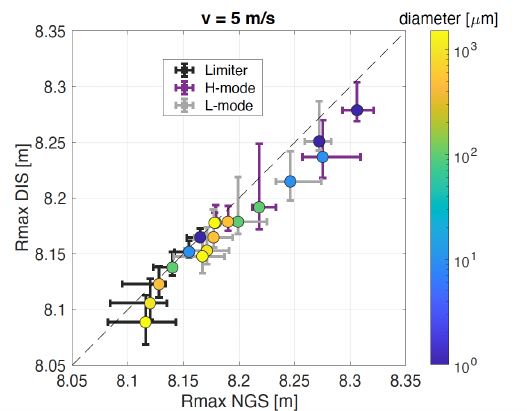}
\caption{Comparison of the radial location of maximum boron evaporation, Rmax, predicted by the DIS and NGS models for particle injection at v=5 m/s across limiter, H-mode, and L-mode ITER scenarios. The color bar indicates the particle diameter and the bars on each point are the Full Width Half Maximum (FWHM) radial extent of the particle deposition from DIS and NGS.}
\label{fig:4}
\end{figure}

Figure \ref{fig:4} shows a comparison of the radial location of maximum boron evaporation injected at 5 m/s across the size scan from 1 \textmu m -- 1500 \textmu m diameter for the DIS and NGS models for all three ITER scenarios. The color within each point indicates the particle diameter, whereas the color of the bars on each point indicates the ITER scenario. The lengths of the bars in the x and y directions indicate the full-width-half-maximum (FWHM) of the radial extent of the evaporated material for the NGS and DIS models, respectively. The dashed line indicates perfect agreement between the two models. Despite significant differences in the physics between the two models, the overall agreement is quite good across a broad range of particle sizes and ITER plasma conditions. Nonetheless, it is also clear from this graph that the DIS model predicts somewhat deeper penetration than the NGS model. In DIS, the particle heating is governed by a time-dependent surface energy balance based on local plasma conditions so that evaporation begins only after the particle temperature reaches the evaporation threshold. NGS instead assumes a fixed particle surface temperature at the evaporation threshold from the moment the particle enters the SOL. The delayed heating due to finite thermal inertia in DIS dominates the shielding effects in NGS so that the particles in DIS take longer to evaporate and penetrate deeper into the plasma before complete evaporation.

\section{Boron injection velocity scan at fixed diameter of 1 mm}
\label{sec:velocity}

\begin{figure}[htbp]
\centering
\includegraphics[width=\linewidth]{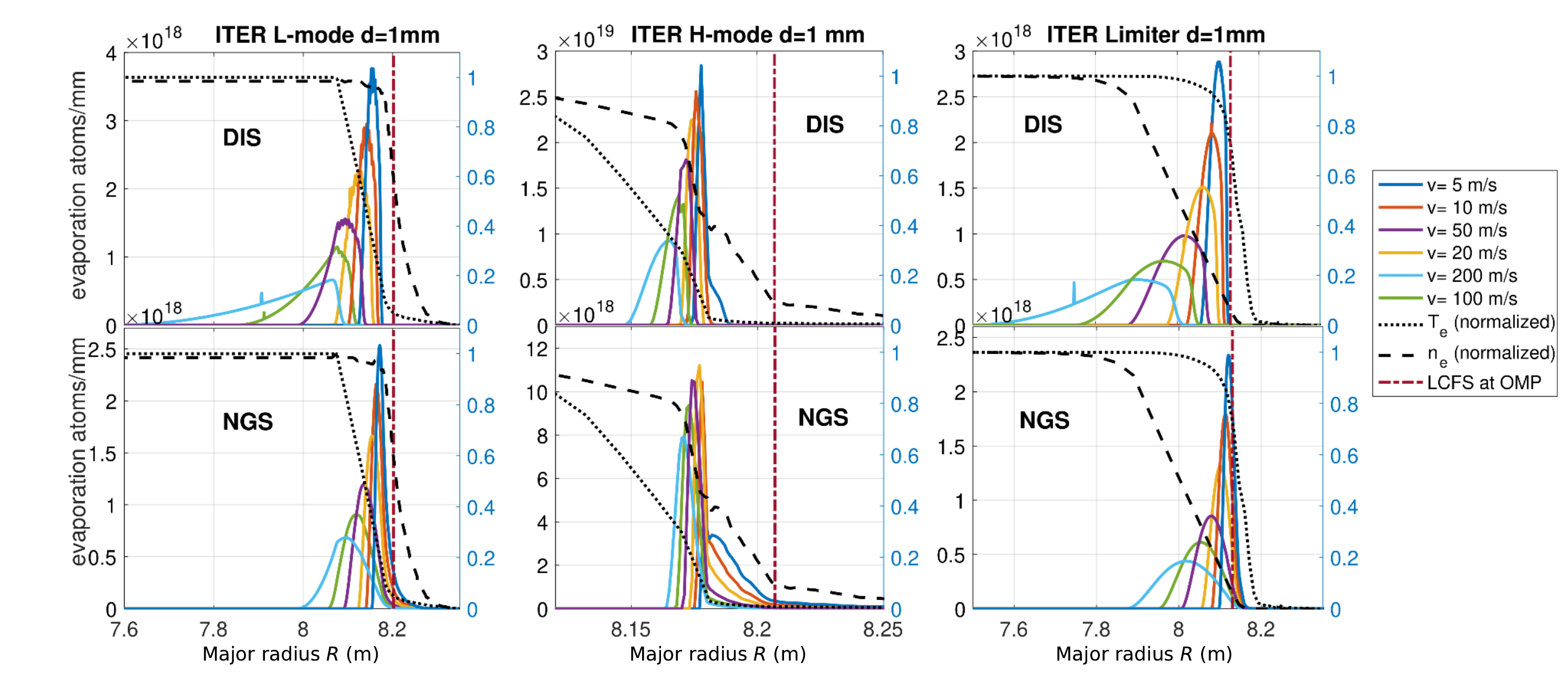}
\caption{Comparison of radial boron evaporation profiles calculated with DIS (top row) and NGS (bottom row) for 1 mm diameter B particles injected at velocities from 5 to 200 m/s in the ITER SRO L-mode (left), Q = 10 H-mode (center), and limiter-startup (right) scenarios. Colored curves show the evaporation rate as a function of major radius, R (m), for the different injection velocities. The normalized background electron temperature and density profiles are overlaid for reference, and the vertical marker indicates the outboard-midplane LCFS position for each scenario.}
\label{fig:5}
\end{figure}

The DIS and NGS models were also compared assuming a fixed diameter of injected boron particles of 1 mm for a velocity scan from 5 m/s to 200 m/s to simulate pellet injection. Figure \ref{fig:5} shows the comparison of the two models across the three ITER scenarios. As expected from the particle size scan at low injection velocity, the 1 mm diameter particles all ablate well inside the LCFS for both models across all three ITER scenarios for all velocities simulated. Comparing the two models, once again, the DIS model shows somewhat deeper penetration than the NGS model, particularly for the limiter and L-mode scenarios. For the Q=10 H-mode scenario, the high local electron temperature and density in the edge pedestal produce a large incident heat flux and rapid ablation in both models and only slightly deeper penetration for the DIS model for the highest velocities.

These results indicate that there is increased penetration with increasing injection velocity in the L-mode and low power limiter cases. The low power limiter case shows significantly deeper penetration well into the plasma as much as 20 -- 30 cm beyond the LCFS at high velocities. In the L-mode case, the particles penetrate 7 -- 15 cm beyond the LCFS for high velocities. In the Q=10 H-mode case, the particles penetrate only about 5 cm beyond the LCFS for all velocities simulated, as the particles reach the high temperature in the edge with little dependence on injection velocity. For both the DIS and NGS models, the particle evaporation rate increases with the incident heat flux, so that the particles quickly evaporate when they hit the high temperature in H-mode, overwhelming the differences between the two models.

\section{Implementation of the boron source and calculation of boron charge-state distributions for future transport modeling}
\label{sec:implementation}

\begin{figure}[htbp]
\centering
\includegraphics[width=0.50\linewidth]{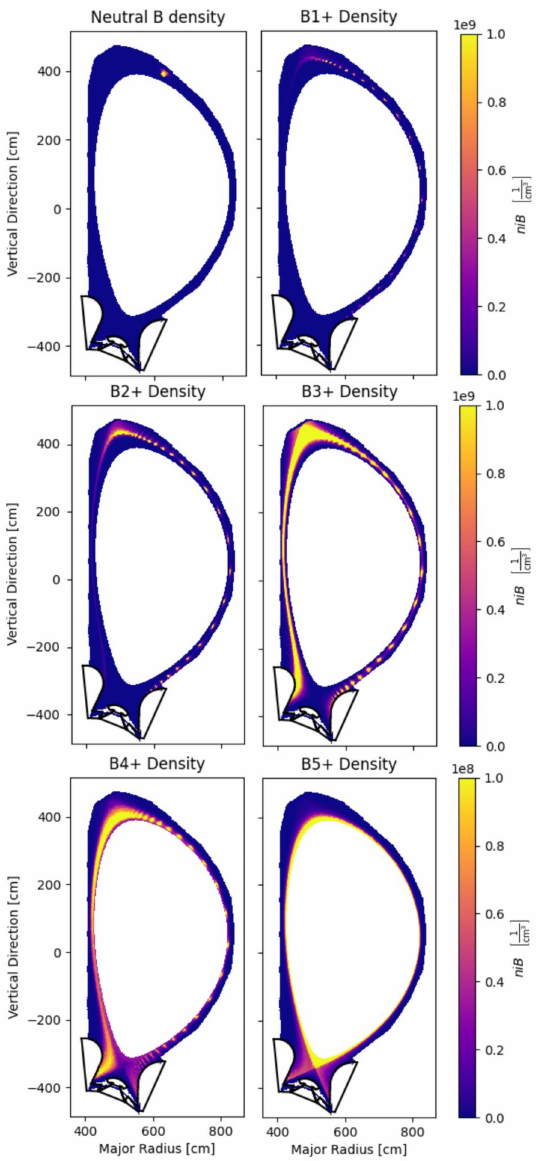}
\caption{Neutral boron source from DIS implemented into EMC3-EIRENE for the ITER Q=10 scenario, showing the calculated spatial distributions of neutral B and the successive boron charge states B\textsuperscript{+1}--B\textsuperscript{+5}. The panels illustrate the ionization sequence and redistribution of injected boron through the edge and scrape-off-layer plasma, from the initial neutral source to progressively higher charge states. Note that the color scale for B\textsuperscript{+4} and B\textsuperscript{+5} differs from that used for the lower charge states.}
\label{fig:6}
\end{figure}

The next step is to implement the boron source term into EMC3-EIRENE to model edge plasma transport and subsequent erosion, deposition, and migration modeling in WallDYN3D \cite{ref24}. Figure \ref{fig:6} shows an initial implementation of the boron source in EMC3-EIRENE for the Q=10 scenario, representative of the source region centered around R=6.15 m and Z=3.95 m associated with the nominal 100 \textmu m, 5 m/s powder-injection condition considered above, together with the five boron charge-state distributions. The neutral source is localized near the upper port injection region. The ionized boron species initially follow magnetic field lines and are then transported toward the divertor from both the inboard and outboard sides. Higher charge states extend progressively farther along the field lines due to rapid ionization in the higher- temperature plasma regions.

These impurity distributions will next be input into WallDYN3D to evaluate migration, erosion, and redeposition to assess surface coatings in ITER, which will be published in a subsequent paper.

\section{Summary and future work}
\label{sec:summary}

This work presents the first step in analyzing solid boron injection for ITER by calculating the boron source term through injection either from an upper port using an Impurity Powder Dropper or through one of the existing inboard or outboard pellet guide tubes using an Impurity Pellet Injector. Two models of SBI injection and ablation (DIS and NGS) were compared across a range of particle sizes from very fine powder to reasonably large pellets of 1.5 mm diameter solid boron and injection velocities from gravitational injection to 200 m/s pellet injection. Although the physics models are quite different between these codes, the overall results agreed reasonably well across several ITER scenarios from a moderate power L-mode, a high-power Q=10 H-mode, and a low power and low current startup scenario. In the SRO L-mode scenario, particles \textgreater{} 100 \textmu m diameter penetrate the LCFS, and particles \textgreater{} 500 \textmu m deposit material well inside the confined plasma, with the B source location shifting by up to \textasciitilde40 cm across the particle-size scan. At Q=10, particles \textgreater{} 500 \textmu m diameter may cross the LCFS, but the high temperature in the H-mode edge immediately ablates even the largest particles simulated. In the low power limiter scenario, the weaker edge plasma provides the least screening of the three cases, allowing larger particles to penetrate substantially farther into the confined plasma region. While larger B pellets up to 5 mm diameter could be injected into ITER high-performance plasmas, the purpose here is to simulate realistic continuous real-time boronization that would be less perturbative to the plasma. The present assessment, and experience from mid-sized tokamaks, indicates that particles of approximately 100 \textmu m diameter should not substantially perturb the ITER plasma, while a quantitative upper limit on non-perturbative particle size is not established here. These results clearly show that the deposition depth can be controlled by choosing the particle size and plasma conditions to ensure the desired penetration depth. The boron source term will next be used to model migration, erosion, and redeposition in ITER in a subsequent publication.

\section*{CRediT authorship contribution statement}
\textbf{F. Effenberg}: Conceptualization, Methodology, Investigation, Writing -- Original Draft, Review \& Editing, Supervision, Project Administration. \textbf{J. A. Snipes}: Conceptualization, Methodology, Investigation, Writing -- Original Draft, Review \& Editing, Supervision, Project Administration. \textbf{F. Nespoli}: Software, Investigation, Writing-- review \& editing. \textbf{R. Lunsford}: Software, Investigation, Visualization, Writing-- review \& editing. \textbf{J. VanBlarcum}: Investigation, Software. \textbf{M. Jia}: Investigation, Software. \textbf{H. Frerichs}: Investigation, Software. \textbf{A. Bortolon}: Investigation. \textbf{K. Schmid}: Investigation, Software. \textbf{T. Wauters}: Investigation.

\section*{Declaration of competing interest}
The authors declare that they have no known competing financial interests or personal relationships that could have appeared to influence the work reported in this paper.

\section*{Data availability}
Data is available at Princeton Commons repository at \url{https://doi.org/10.34770/3tar-6x02} .

\section*{Acknowledgements}
This material is based upon work supported by the U.S. Department of Energy, Office of Science, Office of Fusion Energy Sciences under Award DE-AC02-09CH11466 and the ITER Organization through Task Agreement IO/24/TA/4700000244. The United States Government retains a non-exclusive, paid-up, irrevocable, world-wide license to publish or reproduce the published form of this manuscript, or allow others to do so, for United States Government purposes.

\section*{Disclaimer}
This report was prepared as an account of work sponsored by an agency of the United States Government. Neither the United States Government nor any agency thereof, nor any of their employees, makes any warranty, express or implied, or assumes any legal liability or responsibility for the accuracy, completeness, or usefulness of any information, apparatus, product, or process disclosed, or represents that its use would not infringe privately owned rights. Reference herein to any specific commercial product, process, or service by trade name, trademark, manufacturer, or otherwise does not necessarily constitute or imply its endorsement, recommendation, or favoring by the United States Government or any agency thereof. The views and opinions of authors expressed herein do not necessarily state or reflect those of the United States Government or any agency thereof. ITER is a Nuclear Facility INB-174. The views and opinions expressed herein do not necessarily reflect those of the ITER Organization.

\bibliographystyle{elsarticle-num}
\bibliography{references}

\end{document}